\documentclass[superscriptaddress, twocolumn, amsmath, amssymb, aps, prb]{revtex4-2}

\usepackage{graphicx}
\usepackage{dcolumn}
\usepackage{bm}
\usepackage{float}
\usepackage[colorlinks,
            urlcolor=blue,
            linkcolor=blue,
           anchorcolor=blue,
            citecolor=blue]{hyperref}

\begin{document}

\preprint{APS/123-QED}

\title{Nonreciprocal optical metasurface based on spinning cylinders}

\author{Zheng Yang}
\affiliation{Department of Physics, City University of Hong Kong, Tat Chee Avenue, Kowloon, Hong Kong, China}

\author{Wanyue Xiao}
\affiliation{Department of Physics, City University of Hong Kong, Tat Chee Avenue, Kowloon, Hong Kong, China}

\author{Hengzhi Li}
\affiliation{Department of Physics, City University of Hong Kong, Tat Chee Avenue, Kowloon, Hong Kong, China}

\author{Hao Pan}
\affiliation{Department of Physics, City University of Hong Kong, Tat Chee Avenue, Kowloon, Hong Kong, China}

\author{Shubo Wang} 
\email{\textcolor{black}{shubwang@cityu.edu.hk}} 
\affiliation{Department of Physics, City University of Hong Kong, Tat Chee Avenue, Kowloon, Hong Kong, China}

\date{\today}

\begin{abstract}
Optical systems breaking Lorentz reciprocity have attracted broad attention due to their intriguing physics and applications. Nonreciprocal metasurfaces can enable one-way light transmission and reflection with essential applications in optical communication. Conventional nonreciprocal metasurfaces rely on using magneto-optic or nonlinear materials to induce nonreciprocal optical properties. Here, we propose and demonstrate a new mechanism for realizing nonreciprocal metasurfaces based on the relativistic effect of a moving medium. The metasurface is composed of periodic spinning dielectric cylinders located above a dielectric substrate. The spinning motion breaks the time-reversal symmetry and induces bi-anisotropic Tellegen-type response of the meta-atoms. We show that the metasurface can realize both asymmetric and nonreciprocal manipulations of the incident plane wave. The underlying mechanism is attributed to the Sagnac effect associated with the chiral multipole modes of the coupled spinning cylinders. By introducing dielectric pillars to modulate the phase profile, the metasurface can enable nonreciprocal wavefront manipulations. Our work offers a new mechanism for realizing nonreciprocal light manipulation in free space. The proposed metasurface can serve as a platform to explore the interesting physics of nonreciprocal optics, non-Hermitian optics, and topological photonics.
\end{abstract}

\maketitle


\section{\label{sec:level1}INTRODUCTION}

Optical metasurfaces have demonstrated unprecedented performance in manipulating various light properties, including propagation direction, phase, and polarization \cite{1,chen_review_2016,6}. Numerous intriguing optical functionalities have been realized with metasurfaces, such as beam deflection \cite{12,13}, beam focusing \cite{14,16,19}, and holography \cite{20,22,24}, which significantly facilitate the development of applied optics and give rise to many useful optical devices. Traditional optical metasurfaces obey the time-reversal symmetry and Lorentz reciprocity, which results in a symmetric scattering matrix describing the symmetric response of the metasurfaces under the interchange of the input and output \cite{caloz_electromagnetic_2018}. Breaking Lorentz reciprocity can give rise to nontrivial optical properties and intriguing applications that cannot be realized in conventional reciprocal systems, such as optical isolation \cite{26,29,25,30}, optical circulation \cite{31,32}, and full-duplex optical communication \cite{35,33,34}. 

 So far, the implementation of nonreciprocity can be divided into three broad categories based on their physical mechanisms: magneto-optic effect \cite{haldane_possible_2008,36,45}, nonlinearity \cite{38,39,40,guo_nonreciprocal_2022}, and temporal modulations \cite{41,42,43,taravati_nonreciprocal_2017}. Although these mechanisms have successfully been applied to realize nonreciprocal metasurfaces, they have intrinsic limitations hindering their applications in certain scenarios. The magneto-optic effect usually involves bulky magnets that are difficult to integrate into modern photonic systems \cite{47}. Nonlinear optical systems require a high-intensity field and are reciprocal to noise \cite{38}. The temporal modulation approach implies high complexity in design, especially for a spatiotemporal device such as metasurfaces, and the modulation is usually very weak. Therefore, it is of critical importance to explore new mechanisms of optical nonreciprocity. One robust mechanism is to employ the electromagnetic properties of moving medium \cite{48,mazor_nonreciprocal_2019}. Specifically, spinning structures break the time-reversal symmetry and can give rise to nonreciprocal optical properties \cite{49,50}. This mechanism has been applied to achieve strong optical nonreciprocity in both quantum and classical systems \cite{51,25}. The spinning structures feature a relativistic phenomenon known as the Sagnac effect, which can induce the frequency splitting of a pair of chiral modes \cite{49}. The Sagnac effect enables the selective excitation of a chiral mode and its asymmetric coupling with a guided wave channel, which offers a simple and robust mechanism for realizing non-Hermitian phenomena \cite{50}. 
 
In this article, we apply the moving-medium-induced optical nonreciprocity to realize a nonreciprocal metasurface and investigate its properties associated with light transmission and reflection. The metasurface consists of an array of spinning cylinders (i.e., the meta-atoms) located above a dielectric substrate. We show that the spinning motion induces an effective gauge field that breaks the degeneracy of opposite chiral multipole modes supported by the meta-atoms. Consequently, the meta-atoms exhibit bianisotropic Tellegen-type properties. In addition, the coupling between the meta-atoms can be strongly enhanced by the guided mode supported in the dielectric substrate. Using full-wave numerical simulations and Fourier analysis, we demonstrate strong asymmetric and nonreciprocal light transmission and reflection induced by the metasurface. By introducing an additional degree of freedom to the meta-atoms to control the phase, we further demonstrate nonreciprocal wavefront manipulation with the metasurface.

The airticle is organized as follows. In Sec. \ref{sec:level2}, we discuss the eigenmode properties of the spinning cylinders and the origin of the optical nonreciprocity. In Sec. \ref{sec:level3}, we present and discuss the numerical results for light manipulations enabled by the metasurface composed of spinning cylinders, including asymmetric reflection and transmission, nonreciprocal transmission, and nonreciprocal wavefront manipulation. We draw the conclusion in Sec. \ref{sec:level4}.

\section{\label{sec:level2} Eigenmodes of the SPINNING cylinders}

\begin{figure}[t!]
\centering
\includegraphics[width=0.9\linewidth]{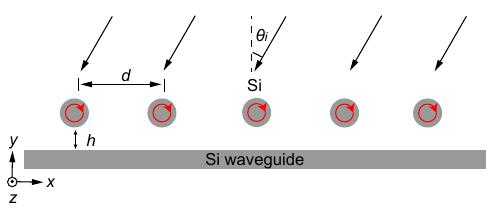}
\caption{Schematic of the nonreciprocal metasurface composed of periodic spinning cylinders placed near a silicon slab with a spacing $h=50$ nm and a period $d=675$ nm. The cylinders have radii $R=200$ nm and rotate at an angular velocity $\Omega$. } \label{Fig1}
\end{figure}

We consider a metasurface composed of silicon cylinders located above a silicon substrate, forming a one-dimensional lattice with period $d=675$ nm, as shown in Fig. \ref{Fig1}. The gap distance between the cylinders and the substrate is $h=50$ nm.  The cylinders have radii $R=200$ nm and spin at angular velocity $\boldsymbol{\Omega}=\Omega \mathbf{\hat{z}}$, where $c$ is the speed of light in vacuum. The whole structure extends infinitely in the $z$ direction. Silicon has relative permittivity $\varepsilon=11.9$ and relative permeability $\mu=1$.

The spinning motion of the cylinders breaks the time-reversal symmetry and turns the cylinders into bianisotropic media \cite{25,30}. The electromagnetic properties of the spinning cylinders can be described by the Minkowski constitutive relations \cite{52}:
\begin{equation}
\begin{aligned}
&\begin{aligned}
& \mathbf{D}+\mathbf{v} \times \frac{\mathbf{H}}{c^2}=\boldsymbol{\varepsilon} \cdot(\mathbf{E}+\mathbf{v} \times \mathbf{B}), \\
& \mathbf{B}+\mathbf{E} \times \frac{\mathbf{v}}{c^2}=\boldsymbol{\mu} \cdot(\mathbf{H}+\mathbf{D} \times \mathbf{v}),
\label{eq:1}
\end{aligned}\\
\end{aligned}
\end{equation}
where $\mathbf{E, D, B}, \text{and }\mathbf{H}$ are the electromagnetic fields; $v(r)=\Omega r$ is the linear speed of an arbitrary point of the cylinders respect to their geometric centers; $\boldsymbol{\varepsilon}$ and $\boldsymbol{\mu}$ are the permittivity and permeability tensors, respectively. Equation (1) indicates that the spinning motion of the cylinders induces the coupling between the electric and magnetic fields. We can rewrite the constitutive relations of Eq.~(\ref{eq:1}) in the matrix form as: 
\begin{equation}
\left[\begin{array}{c}
\mathbf{D} \\
\mathbf{B}
\end{array}\right]=\left[\begin{array}{cc}
\boldsymbol{\varepsilon}^{\prime} & \boldsymbol{\chi}_{\mathrm{em}} \\
\boldsymbol{\chi}_{\mathrm{me}} & \boldsymbol{\mu}^{\prime}
\end{array}\right]\left[\begin{array}{c}
\mathbf{E} \\
\mathbf{H}
\label{eq:2}
\end{array}\right].
\end{equation}
Here, the tensors $\boldsymbol{\chi}_{\mathrm{em}}$ and $\boldsymbol{\chi}_{\mathrm{me}}$ satisfy the relationship $\boldsymbol{\chi}_{\mathrm{em}}=\left(\boldsymbol{\chi}_{\mathrm{me}}^*\right)^{\mathrm{T}}=\left(\boldsymbol{\chi}_{\mathrm{me}}\right)^{\mathrm{T}}$ \cite{49}, which corresponds to the Tellegen-type response that breaks time-reversal symmetry and reciprocity \cite{53}. Equation (\ref{eq:2}) indicates that the spinning cylinders can be treated effectively as stationary cylinders with bi-anisotropic properties, although the base material (i.e. silicon) of the spinning cylinders is isotropic and homogeneous. Note that we use $\boldsymbol{\varepsilon}^{\prime}$ and $\boldsymbol{\mu}^{\prime}$ to denote the effective material parameters of the spinning cylinders.

Using the above constitutive relations, the Maxwell equations under transverse magnetic (TM) polarization can be expressed in local cylindrical coordinates as \cite{25}:
\begin{equation}
\left[\begin{array}{ccc}
-\frac{1}{r} \frac{\partial}{\partial \theta}-i k_0 A_\theta & \frac{1}{r} \frac{\partial}{\partial r} r & -i k_0 \mu_{z}^{\prime} \\
i k_0 \varepsilon_{r}^{\prime} & 0 & \frac{1}{r} \frac{\partial}{\partial \theta}+i k_0 A_\theta \\
0 & i k_0 \varepsilon_{\theta}^{\prime} & -\frac{\partial}{\partial r}
\end{array}\right]\left[\begin{array}{c}
E_r \\
E_\theta \\
Z_0H_z
\end{array}\right]=0.
\label{eq:3}
\end{equation}
where $\varepsilon_{r}^{\prime}$, $\varepsilon_{\theta}^{\prime}$, and $\mu_{z}^{\prime}$ are relative permittivity along the radial and azimuthal directions and the relative permeability along the $z$ direction, respectively; $E_r$, $E_\theta$, and $H_z$ are the field components along the radial, azimuthal, and $z$ directions; $A_\theta=\Omega r\left(1- \varepsilon \mu\right) /\left(\varepsilon \mu r^2\Omega^2/c-c\right)$. We can further obtain the Helmholtz equation by eliminating $E_r$ and $E_\theta$:
\begin{equation}
\left(\frac{1}{r} \frac{\partial}{\partial \theta}+i k_0 A_\theta\right)^2 H_z+\frac{\varepsilon_{r}^{\prime}}{\varepsilon_{\theta}^{\prime}} \frac{1}{r} \frac{\partial}{\partial r}\left(r \frac{\partial}{\partial r} H_z\right)+k_0^2 \varepsilon_{r}^{\prime} \mu_{z}^{\prime} H_z=0 .
\label{eq:4}
\end{equation}
Compared with the Helmholtz equation of a stationary cylinder made of an isotropic material, Eq. (\ref{eq:4}) contains an additional term $i k_0 A_\theta$, which can be considered an effective gauge field in the azimuthal direction and is attributed to the spinning motion of the cylinders. 

The eigenmodes of the cylinders, i.e., the solutions to Eq. (\ref{eq:4}), can be expressed as $H_z=H_z(r) e^{ \pm i m \theta}=H_z(r) e^{ \pm i k_\theta r \theta}$, where $m$ is the azimuthal quantum number of the modes. Substitute the eigenmodes into Eq. (\ref{eq:4}), we find the eigenfrequencies of the modes $H_z(r) e^{ + i m \theta}$ and $H_z(r) e^{ - i m \theta}$  have a difference (i.e., frequency splitting) 
\begin{equation}
\Delta \omega=\frac{2 k_\theta A_\theta c}{A_\theta{ }^2-\varepsilon_{r}^{\prime} \mu_{z}^{\prime}},
\label{eq:5}
\end{equation}
which can be considered a photonic analogue of the Zeeman effect. The above relation is valid under the condition that the linear speed  $\Omega R$  is much smaller than the speed of light $c$ (e.g., $\Omega R/c=0.01$). Under this condition, the higher-order terms in Eq.~(\ref{eq:5}) can be neglected, which leads to the first-order approximation for the frequency splitting  $\Delta \omega=2 m\Omega (\varepsilon \mu-1)/{\varepsilon \mu}$. Based on this expression, we understand that the phenomenon can be enhanced by increasing the mode number $m$ and the spinning speed $\Omega$. For $m\neq0$, the eigenmodes are chiral multipole modes carrying angular momentum. The chiral modes with $m>0$ and $m<0$ rotate in the clockwise (CW) and counterclockwise (CCW) directions, respectively. When the spinning speed $\Omega=0$, corresponding to a stationary cylinder, the modes with $\pm m$ are degenerate and orthogonal due to the mirror symmetry of the cylinder. At a finite spinning speed ($\Omega\neq0$), the mirror symmetry is broken, and the two modes with $\pm m$ are no longer degenerate, corresponding to the frequency splitting of the CW and CCW modes \cite{25,30}. This phenomenon can be understood as the Sagnac effect, which has essential applications such as optical gyroscopes \cite{54}. 

In the presence of the silicon substrate, as shown in Fig. \ref{Fig1}, the spinning cylinders will couple with each other through the guided wave in the substrate. The coupling is generally asymmetric, especially near the resonance frequencies of the CW and CCW modes. This can be understood as follows. The evanescent fields of the guided wave carry transverse spin angular momentum in the $z$ direction. The sign of the transverse spin is locked to the propagating direction of the guided wave, i.e., the guided waves propagating in opposite directions carry opposite transverse spin, a property known as the spin-momentum locking \cite{56,57}. Meanwhile, the CW and CCW chiral modes also carry spin angular momentum in the $+z$ and $-z$ directions, respectively. Due to the frequency splitting property, at a particular frequency, one chiral mode will dominate, and its spin matches the spin of the guided wave propagating in only one direction. Consequently, the couplings among the cylinders are highly asymmetrical near the resonance frequencies of the chiral modes. In the following, we will show that the asymmetrical couplings of the spinning cylinders can give rise to asymmetric reflection and transmission of an incident plane wave. In addition, the asymmetrical couplings play an essential role in inducing strong nonreciprocal properties of the metasurface. 

\begin{figure}[htb]
\centering
\includegraphics[width=0.9\linewidth]{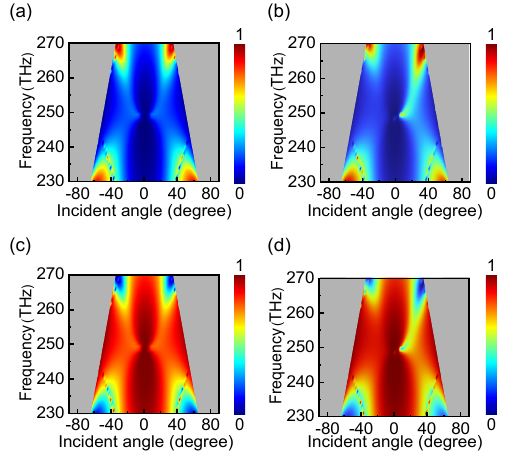}
\caption{Transmission as a function of the frequency and incident angle for the normalized spinning speed (a) $\Omega R/c=0$ and (b) $\Omega R/c=0.01$. Reflection as a function of the frequency and incident angle for the normalized spinning speed (c) $\Omega R/c=0$ and (d) $\Omega R/c=0.01$. The gray areas correspond to the parameter ranges with higher-order diffractions.} \label{Fig2}
\end{figure}

\section{\label{sec:level3} Light manipulations by the metasurface}
\subsection{Asymmetry and nonreciprocity in transmission and reflection}
We consider the metasurface structure in Fig. \ref{Fig1} under the incidence of a linearly polarized plane wave with the magnetic field polarized in the $z$ direction (i.e., TM polarization). The incident angle is denoted by $\theta_i$. The frequency of the plane wave is within [230 THz, 270 THz], corresponding to the subwavelength regime (i.e., $d<\lambda$). Using a finite-element method package COMSOL Multiphysics, we calculate the reflection and transmission of the incident light induced by the metasurface for different incident angles and excitation frequencies. The transmissions are shown in Fig. \ref{Fig2}(a) and \ref{Fig2}(b) for the cases with $\Omega R/c=0$ and $\Omega R/c=0.01$, respectively. The gray regions denote the parameter ranges with higher-order diffraction, which are not of our interest in this work. As seen in Fig. \ref{Fig2}(a), when the cylinders are stationary, the transmission spectrum is symmetric with respect to $\pm \theta_i$. In contrast, the transmission spectrum is obviously asymmetric when the cylinders are spinning, as shown in Fig. \ref{Fig2}(b), which can be attributed to the breaking of the mirror symmetry by the spinning motion of the cylinders. This property also exists in the reflection spectra, as shown in Figs. \ref{Fig2} (c) and \ref{Fig2} (d). The reflection spectra have similar patterns as the transmission spectra since $R=1-T$ under the lossless condition. Importantly, since the plane waves with opposite incident angles $\pm \theta_i$ form a time-reversal pair, the asymmetric reflection spectrum in  Fig. \ref{Fig2}(d) indicates that the metasurface is strongly nonreciprocal. Thus, the nonreciprocity of reflection is directly related to the asymmetry of transmission for the metasurface.

\begin{figure}[tb]
\centering
\includegraphics[width=0.72\linewidth]{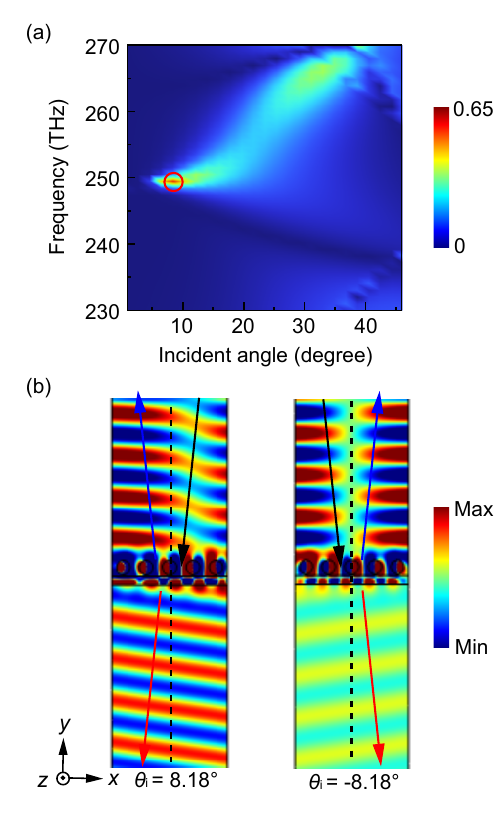}
\caption{(a) Transmission difference (equivalent to reflection difference) as a function of the frequency and incident angle. The red circle marks the maximum transmission difference. (b) Magnetic field $H_z$ under the forward incidence with the incident angle $\theta_i=8.18^{\circ}$ and $\theta_i=-8.18^{\circ}$. The black, red, and blue arrows correspond to the incident, transmitted, and reflected directions, respectively.} \label{Fig3}
\end{figure}

To characterize the asymmetry and nonreciprocity associated with transmission and reflection, we plot the transmission contrast $|T(\theta_i)-T(-\theta_i)|$ in Fig. \ref{Fig3}(a) for the metasurface with $\Omega R/c=0.01$. The transmission contrast is equal to the reflection contrast $|R(\theta_i)-R(-\theta_i)|$ characterizing the nonreciprocity of reflection. We notice that the transmission contrast reaches a maximum value of $65.4 \%$ for the opposite incident angles $\theta_i =\pm 8.18^{\circ}$ at the frequency $f = 249.56$ THz (marked by the red circle). The corresponding magnetic field patterns of the system are shown in Figs. \ref{Fig3}(b) and \ref{Fig3}(c) for $\theta_i=8.18^{\circ}$ and $\theta_i=-8.18^{\circ}$, respectively. The incident, reflected, and transmitted directions are denoted by the black, blue, and red arrows, respectively. In the case of $\theta_i=8.18^{\circ}$, we observe that most of the incident light can transmit through the metasurface and only a small part is reflected. In contrast, in the case of $\theta_i=-8.18^{\circ}$,  most of the incident light is reflected by the metasurface. This demonstrates the strong asymmetry of transmission under opposite incidences $\theta_i =\pm 8.18^{\circ}$ and the strong nonreciprocity of reflection, which can be applied to realize light filtering based on the incident angle.  

\begin{figure}[tb]
\centering
\includegraphics[width=0.68\linewidth]{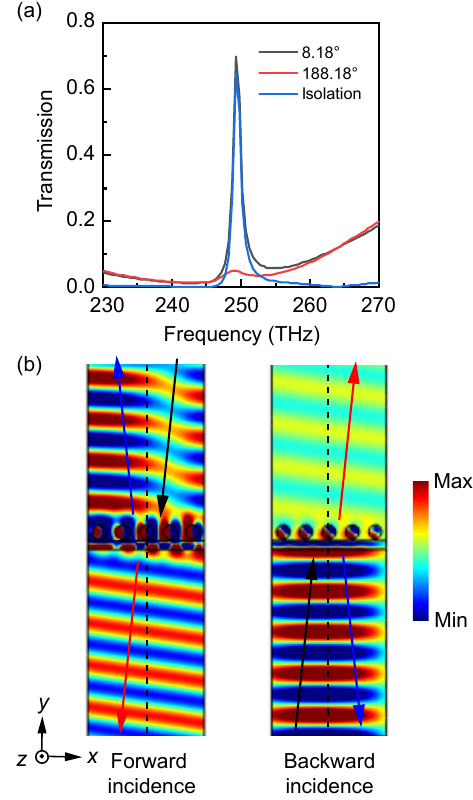}
\caption{(a) Transmission of the metasurface under the forward (solid black line) and backward (solid red line) incidences. The solid blue line denotes the transmission contrast under opposite incidences, corresponding to the isolation ratio. (b) Magnetic field $H_z$ under the forward incidence and backward incidence. The frequency is $f = 249.56$ THz.} \label{Fig4}
\end{figure}

The metasurface in Fig. \ref{Fig1} can also induce nonreciprocal transmission of light. To demonstrate this, we simulate the transmission of the plane waves with incident angles $\theta_i=8.18^{\circ}$ and $\theta_i=188.18^{\circ}$, which correspond to a time-reversal pair of the forward and backward incidences. The numerical results are shown in Fig. \ref{Fig4}(a) as the solid black line and red line, respectively. The blue line denotes the contrast between the forward and backward incidences (i.e., isolation ratio), which characterizes the nonreciprocity of the transmission. We see that the isolation ratio reaches about $65 \%$ at the frequency $f = 249.56$ THz [i.e., the same frequency at which the asymmetry and nonreciprocity of reflection reach the maxima in Fig. \ref{Fig3}(a)]. The same maximum value for the isolation ratio in Fig. 4(a) and the transmission difference in Fig. \ref{Fig3}(a) indicates that the metasurface exhibits similar transmission properties for the mirror-symmetric incidences $\theta_i=-8.18^{\circ}$ and $\theta_i=188.18^{\circ}$. We plot the magnetic field $H_z$ patterns at the resonance frequency in Fig. \ref{Fig4}(b) and \ref{Fig4}(c) for the forward ($\theta_i=8.18^{\circ}$) and backward incidence ($\theta_i=188.18^{\circ}$), respectively. It is obvious that almost all the incident light energy can forwardly transmit through the metasurface, but the backward incident light is nearly completely blocked. This demonstrates that strong nonreciprocal transmission of light can be achieved with the proposed metasurface in the subwavelength regime, which can find applications in designing compact nonreciprocal optical devices such as optical isolators.

\begin{figure}[htb]
\centering
\includegraphics[width=0.95\linewidth]{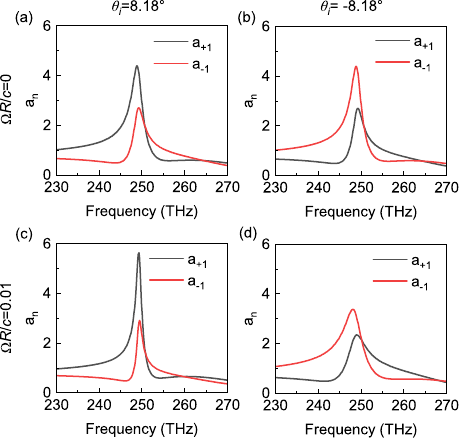}
\caption{Multipole coefficients  $a_{+1}$ and $a_{-1}$ for different incident angles and normalized rotation speeds: (a) $\Omega R/c=0$ and  $\theta_i=8.18^{\circ}$, (b) $\Omega R/c=0$ and  $\theta_i=-8.18^{\circ}$, (c) $\Omega R/c=0.01$ and $\theta_i=8.18^{\circ}$, (d) $\Omega R/c=0.01$ and $\theta_i=-8.18^{\circ}$.} \label{Fig5}
\end{figure}

The strong asymmetric and nonreciprocal properties of light transmission and reflection are closely related to the excitation of the chiral multipole modes supported by the spinning cylinders. To understand this, we conducted multipole expansions for the scattering magnetic field of a spinning cylinder as \cite{bohren_absorption_2008}
\begin{equation}
\mathbf{H}_{\text{s}} =\frac{i}{\omega \mu} \sum_{n=-\infty}^{+\infty} E_{n}\left[b_{n} \mathbf{M}_{n}+i a_{n} \mathbf{N}_{n}\right],
\label{eq:6}
\end{equation}
where $E_n=E_0(-i)^n$ and $\mathbf{M}_n,\mathbf{N}_n$ are the vector harmonics defined in cylindrical coordinate system as:
\begin{equation}
\begin{aligned}
&\mathbf{M}_{n}=i n \frac{Z_{n}(k r)}{r} e^{i n \theta} \hat{r}-k Z_{n}^{\prime}(k r) e^{i n \theta} \hat{\theta}, \\
&\mathbf{N}_{n}=k Z_{n}(k r) e^{i n \theta} \hat{z}.  
\end{aligned} 
\label{eq:7}
\end{equation}
Here, we have $k=k_0,\mu=\mu_0$, $n=m$, and $Z_n=\mathcal{H}_n^{(1)}$ is the Hankel function of the first kind. For the TM polarization in our system with $\mathbf{H}_{\text{s}}=H_z \hat{z}$, only the coefficients $a_n$ exist, which be numerically determined by the integral:
\begin{equation}
\textcolor{black}{a_{n}=\frac{\omega\mu\int_{0}^{2 \pi} \mathbf{N}_{n}^{*} \cdot \mathbf{H}_{s} d \theta}{-E_{n} \int_{0}^{2 \pi} \mathbf{N}_{n}^{*} \cdot \mathbf{N}_{n} d \theta}}, \label{eq:8}
\end{equation}
where “*” denotes the complex conjugate. The integral is carried out on a circle enclosing the spinning cylinder. The obtained coefficients for the dominating modes (i.e., chiral dipole modes $a_{+1}$  and $a_{-1}$) are shown in Fig. \ref{Fig5} for the opposite incident angles $\theta_i=\pm8.18^{\circ}$ and different normalized rotation speeds $\Omega R/c=0$ and $\Omega R/c=0.01$. As seen, under the oblique incidence, the amplitudes of the two chiral dipole modes are generally different due to the breaking of mirror symmetry. For the stationary system, the excitations with opposite incident angles simply lead to a swap of $a_{+1}$  and $a_{-1}$, as shown in Figs. \ref{Fig5}(a) and \ref{Fig5}(b). This is not the case in the spinning system. In the case of Fig. \ref{Fig5}(c) with $\theta_i=8.18^{\circ}$, $a_{+1}$ is much larger than $a_{-1}$ at the resonance frequency. However, in the case of Fig. \ref{Fig5}(d) with $\theta_i=-8.18^{\circ}$, $a_{-1}$ is only slightly larger than $a_{+1}$ at the resonance frequency. This difference in the excitation of the chiral dipole modes gives rise to the asymmetric and nonreciprocal properties of the metasurface. We note that the above multipole expansion only serves as a qualitative explanation since $\mathbf{H}_{\text{s}}$ contains the fields of all cylinders and the substrate. For a quantitative explanation, it is necessary to develop a multipole expansion based on the current inside the cylinder \cite{grahn_electromagnetic_2012}.

\begin{figure}[tb]
\centering
\includegraphics[width=0.75\linewidth]{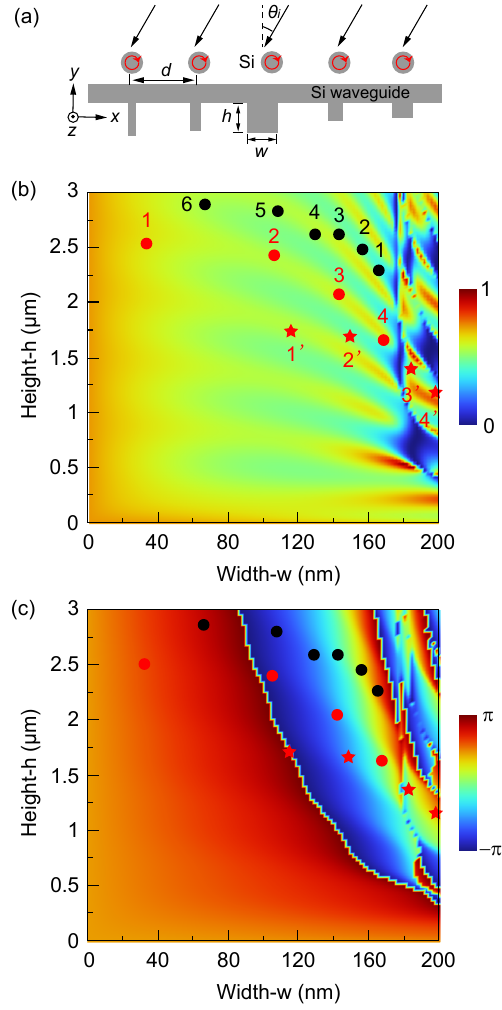}
\caption{(a) Schematic of the metasurface for wavefront manipulation. (b) Amplitude and (c) phase of the transmitted wave when the incident angle is $\theta_i=8.18^{\circ}$ and the frequency is $f = 249.56$ THz. The symbols (dots and stars) mark the parameters chosen for the designed metasurfaces (see text for details).} \label{Fig6}
\end{figure}

\subsection{Nonreciprocal wavefront manipulation}
In recent years, various metasurfaces have been proposed to control optical wavefront, which usually employ three types of phases: resonance phase \cite{60,61}, geometric phase \cite{69,marrucci_optical_2006,70,64}, and propagation phase \cite{66,67}. The resonance phase can induce higher-order diffractions in strongly coupled periodic systems. The geometric phase is usually exploited for circularly polarized incident light. The propagation phase is general and can be controlled by simply altering the geometric size of the meta-atoms. Here, we control the propagation phase with the metasurface to manipulate the wavefront of incident light while maintaining a strong nonreciprocity of transmission. 

To control the propagation phase, we introduce additional degrees of freedom to the metasurface by adding dielectric pillars to the meta-atoms, as shown in Fig. \ref{Fig6}(a). The material of the pillars is the same as that of the substrate (i.e., silicon). We adjust the height $h$ and the width $w$ of the pillars to tailor the local phase of the transmitted wave. Figures \ref{Fig6}(b) and \ref{Fig6}(c) show the numerically calculated transmission amplitude and phase, respectively, as a function of $h$ and $w$. We set the incident angle $\theta_i=8.18^{\circ}$, the frequency $f = 249.56$ THz, and the spinning speed  $\Omega R/c=0.01$. We notice that the amplitude is approximately a constant when the width $w$ is small but strongly varies at large values of $w$ due to Fabry-Perot resonances. Meanwhile, the phase generally varies with $h$ and $w$. Importantly, the phase can cover $2 \pi$ full range. This allows us to choose appropriate pillars to form supercells to modulate the transmission wavefront. In the following, we choose three sets of pillars to realize nonreciprocal beam deflections.  

\begin{figure}[tb]
\centering
\includegraphics[width=0.75\linewidth]{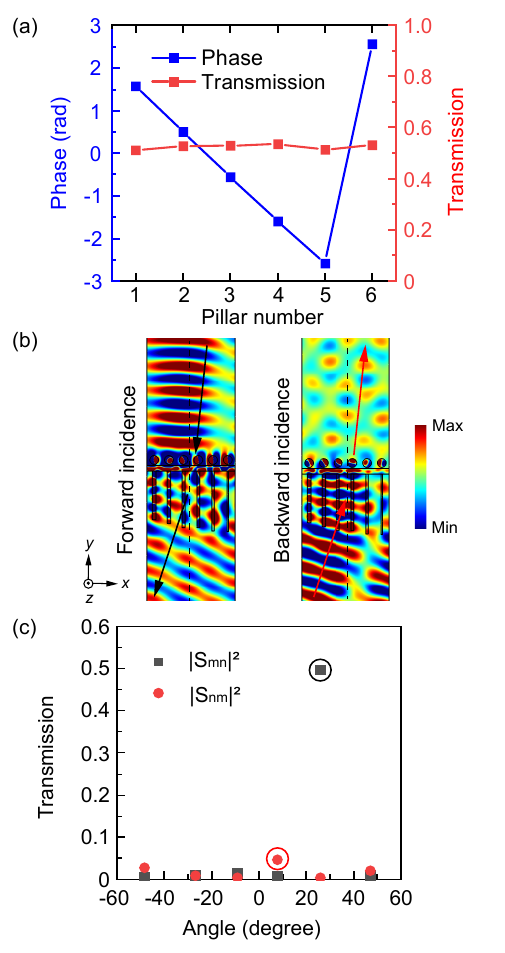}
\caption{(a) Transmission phase and amplitude for the pillars with geometric parameters labeled as No. 1-6 in black in Fig. \ref{Fig6}. (b) Magnetic field $H_z$ for the forward incidence (left panel) and backward incidence (right panel). (c) Fourier analysis for the amplitudes of the transmission channels under forward incidence (black dots) and backward incidence (red dots). The dots surrounded by black and red circles are the channels $|S_{21} |^2$ and $|S_{12} |^2$.} \label{Fig7}
\end{figure}

The first set of pillars have the width $w$ and height $h$ marked by the black dots in Figs. \ref{Fig6}(b) and \ref{Fig6}(c), which are labeled as No. 1-6. Figure \ref{Fig7}(a) shows the transmission amplitude (red dotted line) and phase (blue dotted line) corresponding to the six pillars. The pillars No. 1-6 are placed on the substrate from left to right to form a supercell with periodic boundary condition, giving rise to a phase difference of $\mathrm{d}\phi=-\pi/3$, where $\phi$ is the local phase. We set the frequency $f=249.56$ THz, the period $d=675$ nm, and the incident angle $\theta_i=8.18^{\circ}$. The deflection angle $\theta_t$ of the transmitted wave can be determined by the generalized Snell’s law \cite{68}:
\begin{equation}
\sin \left(\theta_t\right)-\sin \left(\theta_i\right)=-\frac{\lambda}{2 \pi} \frac{\mathrm{d} \phi}{\mathrm{d} x}.
\label{eq:9}
\end{equation}
In this case, the output angle predicted by the above equation is $\theta_t=26^{\circ}$. To demonstrate the nonreciprocal wave steering by the metasurface, we simulate the transmission of a linearly polarized plane wave under the forward and backward incidences. In the case of forward incidence with  $\theta_i=8.18^{\circ}$, the transmitted wave propagates in the direction forming an angle of $\theta_t=26^{\circ}$ with the $y$ axis, as shown in the left panel of Fig. \ref{Fig7}(b), agreeing with the predicted transmission angle. The transmission of power for this case is $53.8 \%$. In the case of backward incidence, the transmitted wave exhibits interference due to the emergence of high-order diffractions, as shown in the right panel of Fig. \ref{Fig7}(b). The transmission of power for this case is $10.8 \%$. 

To determine the strength of nonreciprocity, we conduct Fourier analysis for the transmissions under forward and backward incidences to obtain the contribution of each diffraction channel. The results are summarized in Fig. \ref{Fig7}(c). The black squares denote the transmission channels under forward incidence, which is dominated by the channel at the diffraction angle of $26^{\circ}$ (marked by a black circle) with a contribution of $|S_{21} |^2=49.2 \%$. The red dots denote the transmission channels under the backward incidence. The channel corresponding to the time reversal of the forward incidence (at the diffraction angle $8.18^{\circ}$) is marked by the red circle, which contributes $|S_{12} |^2=4.4 \%$ to the total transmission. Thus, the isolation ratio is $||S_{21} |^2-|S_{12} |^2 |=44.8 \%$, which is smaller than in the case of Fig. \ref{Fig4} due to the high order diffractions. 

\begin{figure}[tb]
\centering
\includegraphics[width=0.75\linewidth]{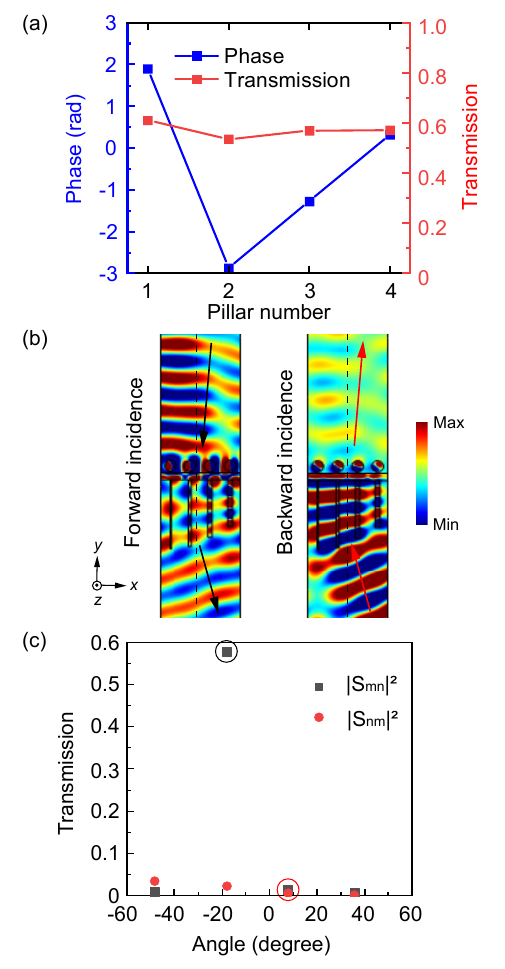}
\caption{(a) Transmission phase and amplitude for the pillars with geometric parameters labeled as No. 1-4 in red in Fig. \ref{Fig6}. (b) Magnetic field $H_z$ for the forward incidence (left panel) and backward incidence (right panel). (c) Fourier analysis for the amplitudes of the transmission channels under forward incidence (black dots) and backward incidence (red dots). The dots surrounded by black and red circles are the channels $|S_{21} |^2$ and $|S_{12} |^2$.}. \label{Fig8}
\end{figure}

To demonstrate the robustness of the mechanism, we choose another set of pillars with the geometric sizes marked by the red dots in Fig. \ref{Fig6}(b) and \ref{Fig6}(c), which are labeled as No. 1-4. Figure \ref{Fig8}(a) shows the numerically calculated transmission amplitude (red dotted line) and phase (blue dotted line) corresponding to the four pillars. The pillars are placed on the substrate from left to right to form the supercell with periodic boundary condition, giving rise to a phase difference $\mathrm{d}\phi=\pi/2$. According to Eq.~(\ref{eq:9}), the deflection angle should be $\theta_t=-17.5^{\circ}$ for this case with an incident angle of $\theta_i=8.18^{\circ}$, i.e., negative deflection can be achieved with this metasurface. The phenomenon is confirmed by full-wave simulations, as shown in Fig. \ref{Fig8}(b). We notice that the plane wave is negatively deflected under the forward incidence but is nearly blocked under the backward incidence. Similar to the case of Fig. \ref{Fig7}, we conduct Fourier analysis of the transmissions and obtain the contributions of different diffraction channels. The results are summarized in Fig. \ref{Fig8}(c). The black square and red dot marked by the circles correspond to the transmissions of a time-reversal pair under opposite incidences, which have $|S_{21} |^2=57.7 \%$ and $|S_{12} |^2=0.6 \%$, respectively. Thus, the isolation ratio is $||S_{21} |^2-|S_{12} |^2 |=57.1 \%$, which is larger than that of the case in Fig. \ref{Fig7}. This is attributed to a higher transmission of the metasurface with four pillars.

\begin{figure}[tb]
\centering
\includegraphics[width=0.75\linewidth]{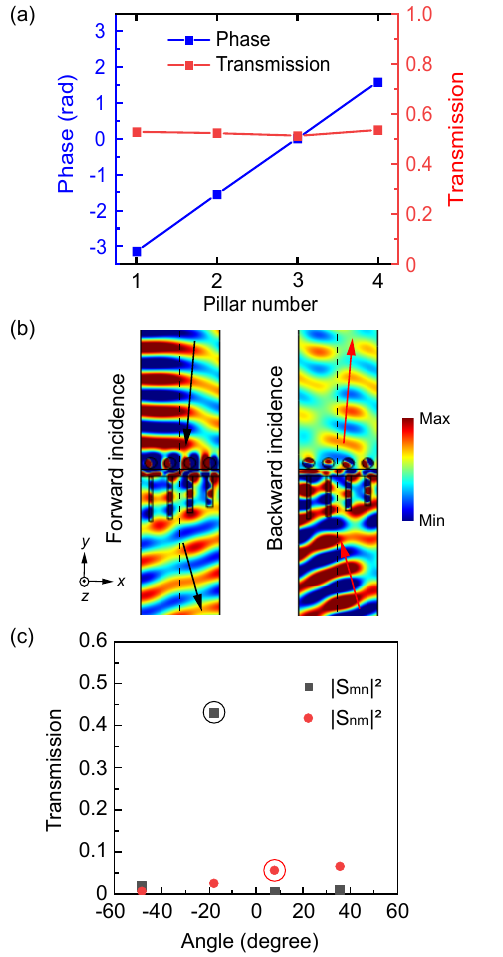}
\caption{(a) Transmission phase and amplitude for the pillars with geometric parameters labeled as No. 1’-4’ in red in Fig. \ref{Fig6}. (b) Magnetic field $H_z$ for the forward incidence (left panel) and backward incidence (right panel). (c) Fourier analysis for the amplitudes of the transmission channels under forward incidence (black dots) and backward incidence (red dots). The dots surrounded by black and red circles are the channels $|S_{21} |^2$ and $|S_{12} |^2$.}. \label{Fig9}
\end{figure}

The above two metasurfaces have thickness larger than the working wavelength due to the utilization of the propagation phase. To further demonstrate the robustness of the mechanism, we present a third metasurface with an overall thickness of 1700 nm, which is comparable to the working wavelength $\lambda=1202$ nm. The geometric dimensions of the pillars are marked by the four red stars in Fig. \ref{Fig6}(a) and \ref{Fig6}(b) and labeled as No. 1’-4’. The corresponding transmission amplitude and phase are shown in Fig. \ref{Fig9}(a). The incident and deflection angles in this case are $8.18^{\circ}$ and $-17.5^{\circ}$, respectively. The numerically simulated field patterns for the forward and backward incidence are shown in Fig. \ref{Fig9}(b), which are similar to the four-pillar case in Fig. \ref{Fig8}. The Fourier analysis of the transmissions is shown in Fig. \ref{Fig9}(c), where black square and red dot with circles correspond to the transmissions of a time-reversal pair, which have values $|S_{21} |^2=43 \%$ and $|S_{12} |^2=5.4 \%$, respectively. The isolation ratio for this case is $||S_{21} |^2-|S_{12} |^2 |=37.6 \%$. This metasurface has a smaller thickness but weaker nonreciprocity compared to the metasurface in Fig. \ref{Fig8}, which may be attributed to the Fabry-Perot resonances of the pillars with larger values of $w$.

\section{\label{sec:level4}CONCLUSION}
In summary, we propose a new mechanism to realize nonreciprocal metasurfaces based on the relativistic effect of spinning dielectric cylinders coupled through a dielectric substrate. The spinning motion breaks the time-reversal symmetry and leads to bi-anisotropic Tellegen-type properties of the cylinders. This gives rise to a synthetic gauge field acting on the cylinders to remove the degeneracy of opposite chiral multipole modes. The spinning motion also breaks the mirror symmetry and induces asymmetric reflection and transmission of the incident plane wave. We show that the metasurface can be applied to achieve nonreciprocal reflection and transmission of light. In addition, we demonstrate nonreciprocal wavefront manipulation with the metasurface by introducing dielectric pillars to control the local propagation phase. We realize both positive and negative one-way deflections of the incident linearly polarized plane wave, with the deflection angle decided by the phase gradient induced by the pillars and the isolation ratio mainly decided by the properties of the spinning cylinders. 

Our work uncovers the relativistic effect on electromagnetic nonreciprocity and contributes to the understanding of the nonreciprocal properties of moving media in periodic structures. The results can be applied to design compact nonreciprocal optical devices for communications. The proposed mechanism of nonreciprocity may also find applications in non-Hermitian optics and topological photonics, where the proposed metasurface can serve as a convenient platform to explore the interesting physics derived from the interplay between nonreciprocity, non-Hermiticity, and topology.

\begin{acknowledgments}
The work described in this paper was supported by grants from the Research Grants Council of the Hong Kong Special Administrative Region, China (Projects No. CityU 11301820 and No. AoE/P-502/20) and National Natural Science Foundation of China (No. 12322416).
\end{acknowledgments}

\bibliography{Reference}

\end{document}